\documentclass[12pt]{article}
\usepackage{graphicx}

\newcommand{\PTm}{\ensuremath{{p}_\mathrm{T}\hspace{-1.02em}/\kern 0.5em}\xspace}

\newcommand{\ETslash}{\ensuremath{E_{\mathrm{T}}\hspace{-1.1em}/\kern0.45em}\xspace}

\newcommand{\ptvecmiss}{\ensuremath{{\vec p}_{\mathrm{T}}^{\kern1pt\text{miss}}}\xspace}

\newcommand{\HGG}{\ensuremath{\mathrm{H}\to\gamma\gamma}}

\newcommand{\rpv}{\ensuremath{\rlap{\kern.2em/}R}\xspace}

\def\Title#1{\begin{center} {\Large #1 } \end{center}}
\def\Author#1{\begin{center}{ \sc #1} \end{center}}
\def\Ead#1{\begin{center}{ #1} \end{center}}
\def\Address#1{\begin{center}{ \it #1} \end{center}}

\newenvironment{Abstract}{\begin{quotation}  }{\end{quotation}}
\newenvironment{Presented}{\begin{quotation} \begin{center} 
             PRESENTED AT\end{center}\bigskip 
      \begin{center}\begin{large}}{\end{large}\end{center} \end{quotation}}

\def\beq{\begin{equation}}
\def\eeq#1{\label{#1}\end{equation}}
\def\eeqn{\end{equation}}

\def\beqa{\begin{eqnarray}}
\def\eeqa#1{\label{#1}\end{eqnarray}}
\def\eeqan{\end{eqnarray}}

\let\bar=\overbar

\def\Dslash{\not{\hbox{\kern-4pt $D$}}}
\def\dslash{\not{\hbox{\kern-2pt $\del$}}}

\def\msb{{\bar{\ssstyle M \kern -1pt S}}}

\begin{document}
\begin{titlepage}

\vfill
\Title{Precision Timing with the CMS MIP detector}
\vfill
\Author{Irene Dutta on behalf of the CMS collaboration}
\Address{California Institute of Technology, Pasadena, CA, USA - 91125}
\Ead{Email : idutta@caltech.edu}
\vfill
\begin{Abstract}
The Compact Muon Solenoid (CMS) detector at the CERN Large Hadron Collider (LHC) is undergoing an extensive Phase II upgrade program to prepare for the challenging conditions of the High-Luminosity LHC (HL-LHC). A new timing layer is designed to measure minimum ionizing particles (MIPs) with a time resolution of $\sim$30 ps and a hermetic coverage up to a pseudo-rapidity of $|\eta|$ = 3. This MIP Timing Detector (MTD) will consist of a central barrel region based on LYSO:Ce crystals read out with SiPMs and two end-caps instrumented with radiation-tolerant Low Gain Avalanche Diodes (LGADs). The precision time information from the MTD will reduce the effects of the high levels of pile-up expected at the HL-LHC, and will bring new and unique capabilities to the CMS detector. We present the current status and ongoing R$\&$D of the MTD, including recent test beam results.
\end{Abstract}
\vfill
\begin{Presented}
CIPANP 2018\\
Palm Springs, CA, USA,  May 29--June 3, 2018
\end{Presented}
\vfill
\end{titlepage}

\section{Introduction}
The future of the LHC is the High-Luminosity LHC (HL-LHC) ~\cite{Apollinari:2017cqg}, which is scheduled to begin operation by the year 2026.  HL-LHC will operate at a stable luminosity of 5.0$\times 10^{34}$~cm$^{-2}$s$^{-1}$, yielding 140 pileup collisions during a fill. In an ultimate scenario, it plans to operate with 7.5$\times 10^{34}$~cm$^{-2}$s$^{-1}$ luminosity and 200 pileup collisions per beam crossing, in order to achieve its goal of accumulating $\sim 3000fb^{-1}$ of data. This amount of data opens up new unexplored regions of phase space for Beyond Standard Model (BSM) searches while also allowing to make more precise Standard Model (SM) measurements.

The increased amount of pileup also presents new challenges. Current pileup mitigation in CMS relies upon particle-flow event reconstruction~\cite{CMS:2009nxa} that removes charged tracks inconsistent with the main vertex of interest and also neutral deposits in the calorimeters with ansatz-based statistical inference techniques like PUPPI~\cite{Bertolini:2014bba}. However, the increase in the spatial overlap of tracks and energy deposits from these collisions will deteriorate the identification and the reconstruction efficiency of the hard interaction using standard techniques.\\

The Phase-2 upgrades to the CMS detector~\cite{Butler:2020886,Butler:2055167} for the High-Luminosity LHC (HL-LHC) are required to maintain the excellent performance of the detector in terms of efficiency, resolution, and background rejection for all final state particles and physical quantities used in data analyses. These upgrades are necessary to withstand radiation damage effects and overcome the challenges posed by the high rate of concurrent collisions per beam crossing (pileup) at the HL-LHC.\\

To improve the particle-flow performance at high pileup to a level comparable to that of the Phase-1 CMS detector, there will be two upgrades: enhanced timing capabilities of the  calorimeters~\cite{Butler:2020886} and a dedicated detector for precision timing of minimum ionizing particles (MIPs): the MIP timing detector (MTD)~\cite{Collaboration:2296612}. Pileup collisions at the HL-LHC will occur with an RMS spread of approximately 180-200 ps, constant during every bunch crossing. Slicing the beam spot in consecutive 30 ps exposures effectively reduces the number of vertices down to current LHC conditions, thereby recovering the Phase-1 quality of event reconstruction.
\begin{figure}[t!]
\centering
\raisebox{-0.45\height}{\includegraphics[width=0.6\textwidth]{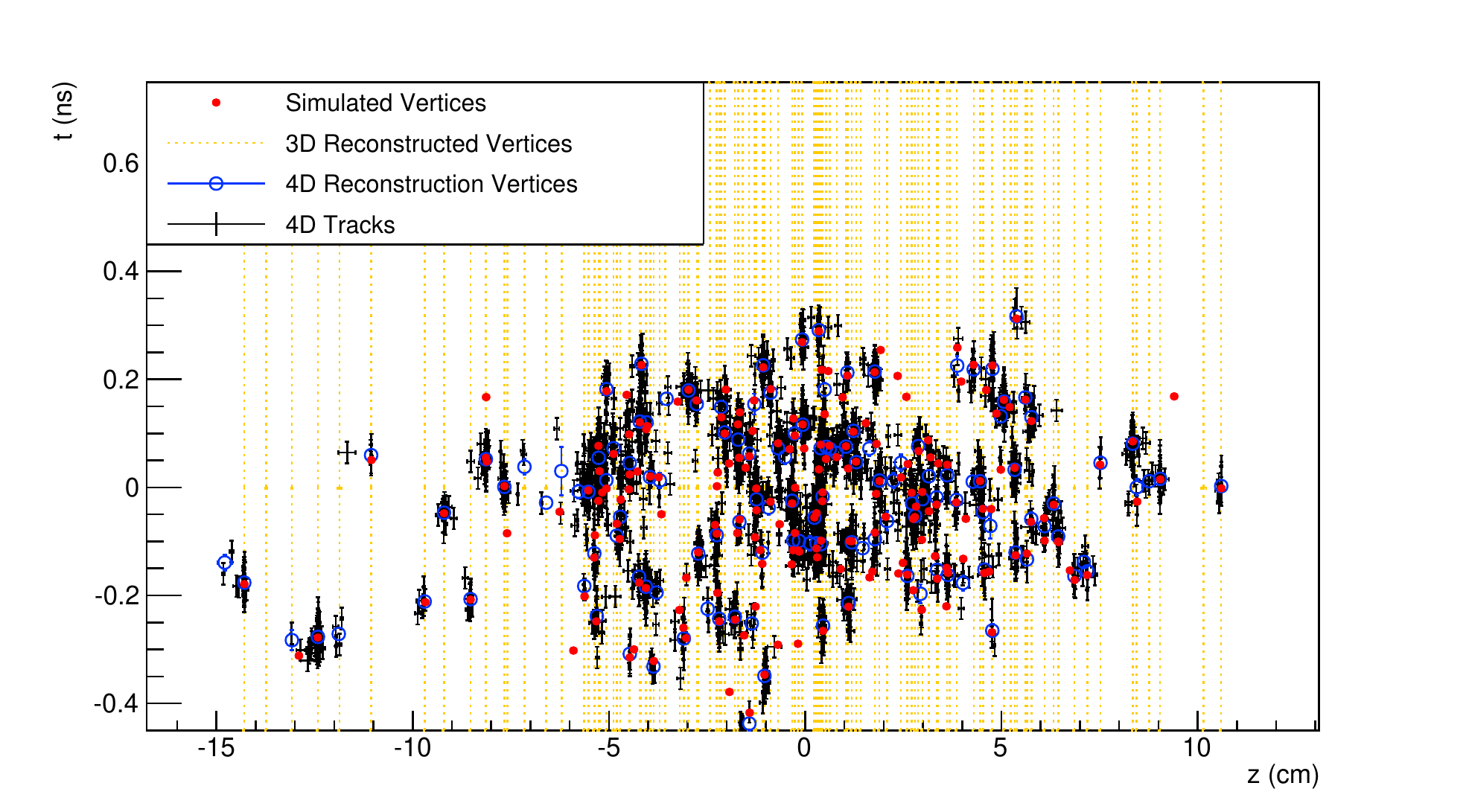}}
\raisebox{-0.5\height}{\includegraphics[width=0.35\textwidth]{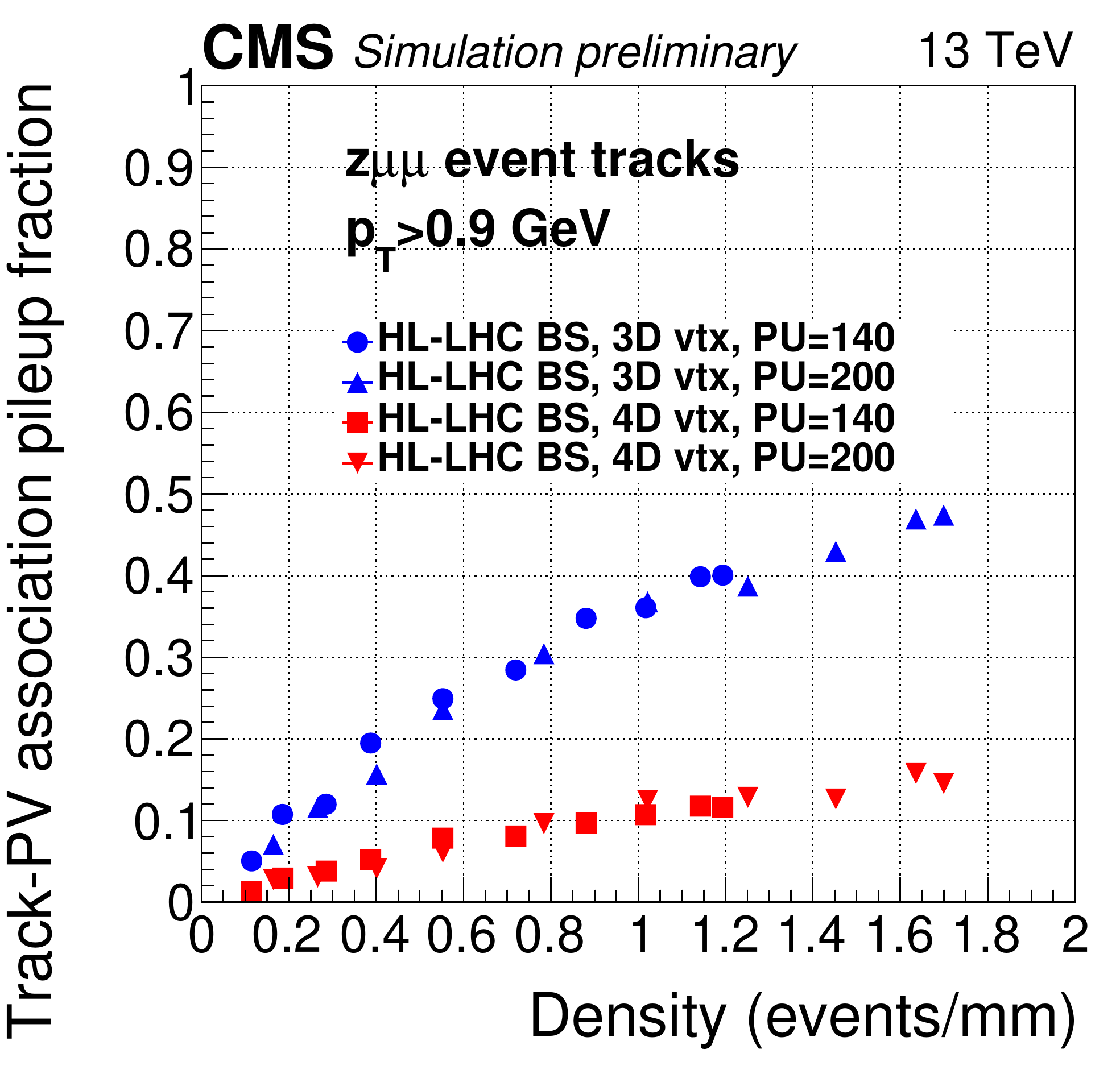}}
\caption{Left: Simulated and reconstructed vertices in a 200 pileup event assuming a MIP timing detector covering the barrel and endcaps. The vertical lines indicate 3D-reconstructed vertices, with instances of vertex merging visible throughout the event display. Right: Rate of tracks from pileup vertices incorrectly associated with the primary vertex of the hard interaction normalized to the total number of tracks in the vertex.
}
\label{fig:4Dvertex_200}
\end{figure}

Fig.~\ref{fig:4Dvertex_200} (Left panel) demonstrates the fact that vertex merging is reduced from 15\% in space to 1\% in space-time with 4D information. The right panel of Fig.~\ref{fig:4Dvertex_200} shows the improvement in correct track-vertex association with the added timing information in the high pileup case.
Timing information can improve the performance of b-jet identification (based on secondary vertex reconstruction) and also improves the isolation efficiency of leptons and photons by removing unwanted pileup tracks. It also helps in the better reconstruction of jets and missing transverse momentum, which are very vulnerable to pileup.

\section{Impact of precision timing on the HL-LHC physics program}
The CMS physics program at the HL-LHC will target a very wide range of measurements, including precise measurements of the Higgs boson properties and direct searches for physics beyond the standard model (BSM). All these studies will benefit from the improved acceptance for isolated objects and in the case of \HGG{} decays, there is an added benefit from improved vertex identification.
\begin{figure}[hbtp]
\centering
\raisebox{-0.5\height}{\includegraphics[width=0.36\textwidth]{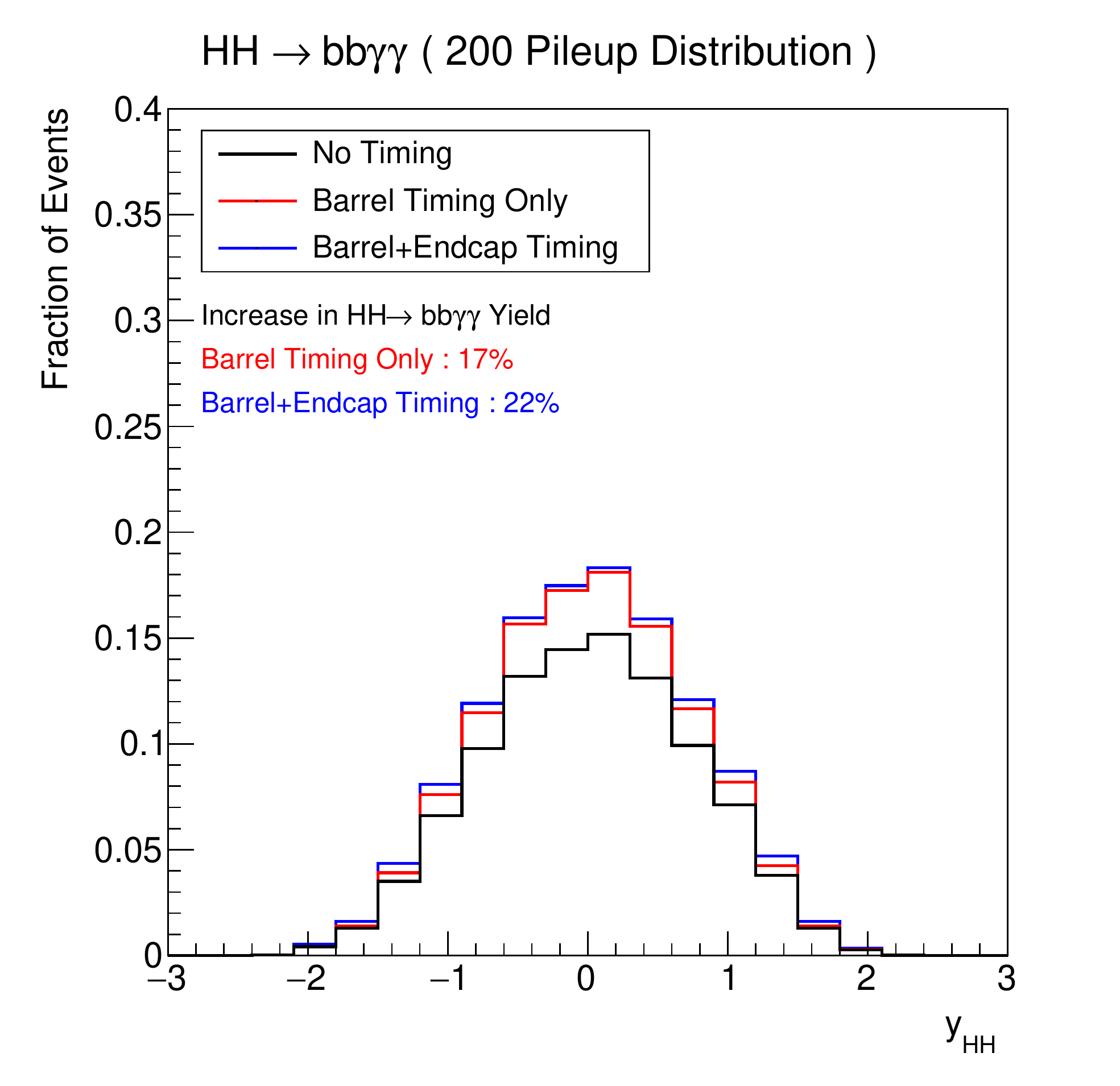}}
\caption{Impact on signal efficiency for HH$ \to b \bar{b} \gamma \gamma$ for no-timing layer vs with a timing layer implemented(left). }
\label{fig:HHAndllpSUSY}
\end{figure}
A very crucial measurement to be performed at the HL-LHC is the di-Higgs production, which is a direct measurement of the Higgs self-coupling. In this case, precision timing increases the signal yields for constant background in HH$ \to b \bar{b} \gamma \gamma$ by 17\% from the barrel alone, and 22\% with hermetic coverage (Fig.~\ref{fig:HHAndllpSUSY}). Similar improvements are predicted for other Higgs boson signatures, ranging from 15--20\% for HH$\to 4$b to 20--26\% for H$\to4\mu$, for a constant background.

The sensitivity to several {searches for new phenomena} is largely driven by the $E_{T}^{miss}$ resolution, which determines the background level for several BSM signatures. The gain in the $E_{T}^{miss}$ resolution with track timing leads to a reduction of $\sim 40$\% in the tail of the $E_{T}^{miss}$ distribution above 130~GeV, which approximately offsets the performance degradation for SUSY searches in the transition from 140 to 200 pileup~\cite{Butler:2055167}.

\section{MIP Timing Detector}
\label{sec:MTDoverview}
\begin{figure*}[t!]
\centering
\includegraphics[width=0.68\textwidth]{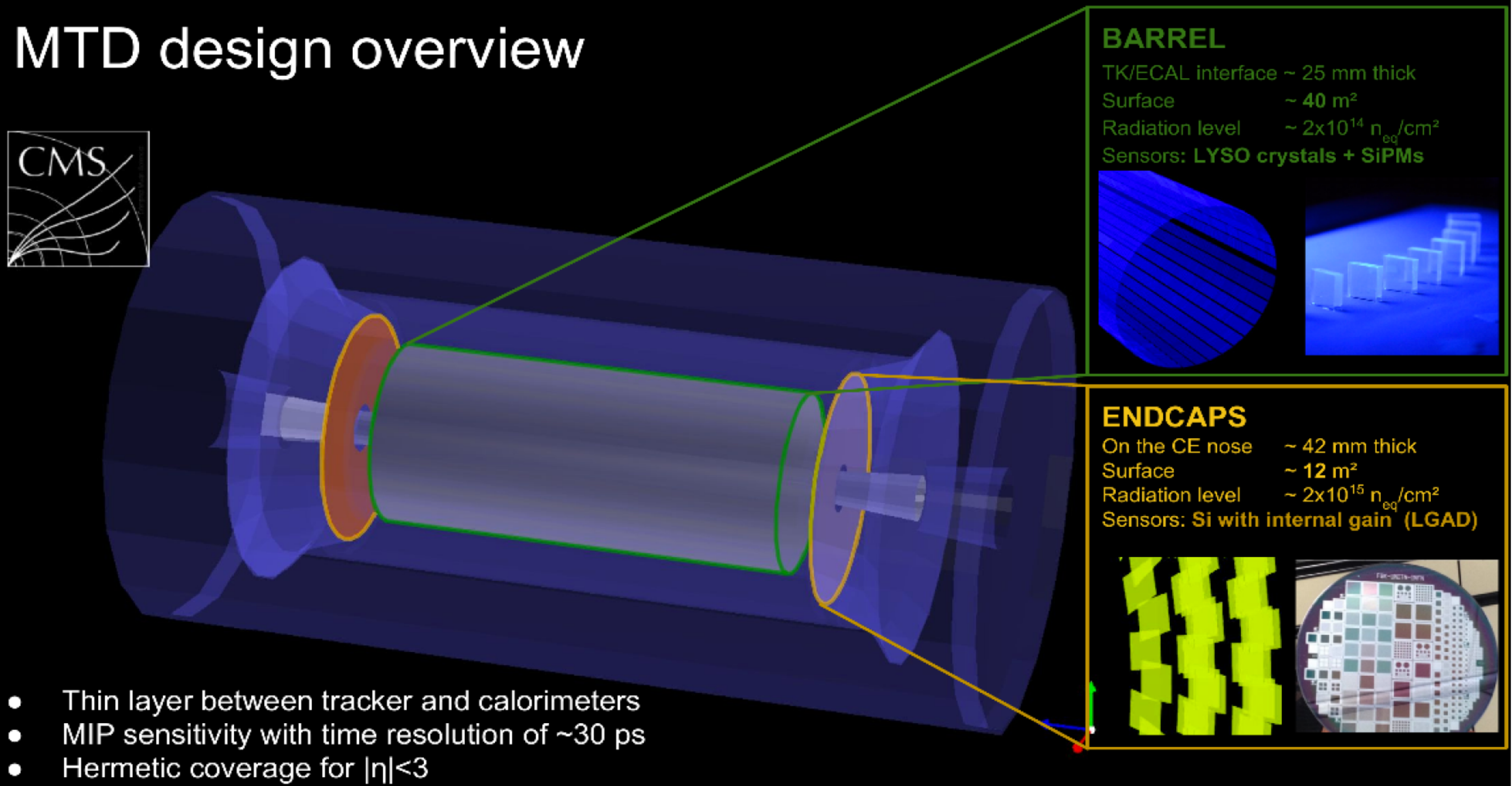}
\caption{
A simplified geometry of the timing layer implemented for simulation studies. The layer will be built at the interface between the Tracker and the ECAL and comprises a SiPM+LYSO:Ce barrel (grey cylinder) and two endcap (orange discs) timing layers in front of the endcap calorimeter.
}
\label{fig:EvtDisplay}
\end{figure*}
Figure~\ref{fig:EvtDisplay} shows a simplified implementation in $\textsc{geant}$ of the proposed layout of the MTD in the CMS detector. The MTD will comprise a barrel and an endcap region, with different technologies based on different requirements, for e.g., radiation dose, schedule constraints, cost effectiveness etc. Some of the key features that led to the design technology are listed in Table~\ref{table:design}.

\begin{table}
\centering
\scriptsize
\caption{Key features of the MTD.}
\label{table:design}
\begin{tabular}{|c|c|c|} \hline
  & Barrel (LYSO:Ce+SiPM) & Endcap (LGAD) \\ \hline
Region  & $|\eta| <$ 1.5 & 1.6$ <|\eta| <$ 3.0  \\ \hline
Surface Area   & $36.5 m^{2}$ 	& $12 m^{2}$  \\ \hline
Power consumption   & 0.5 kW/$m^{2}$ 	& 1.8 kW/$m^{2}$ \\ \hline
Radiation Dose   & 2$\times 10^{14} n_{eq}/cm^{2}$ 	&   2$\times 10^{15} n_{eq}/cm^{2}$\\\hline
Installation Dose & 2022 & 2024 \\
\hline
\end{tabular}
\end{table}

Crystal scintillators read out with silicon photomultipliers (SiPMs)~\cite{gundacker2013time, LYSONIM, Anderson:2015tia} and silicon sensors with internal gain~\cite{White:2014oga,Pellegrini201412, Cartiglia:2015iua} were chosen as a mature technology for the barrel and a viable technology for the endcap timing layers respectively. The following sections discuss these technologies in more detail.

\subsection{Barrel Timing Layer (BTL)}
\begin{figure}[hbtp]
\centering
\raisebox{-0.5\height}{\includegraphics[width=0.34\textwidth]{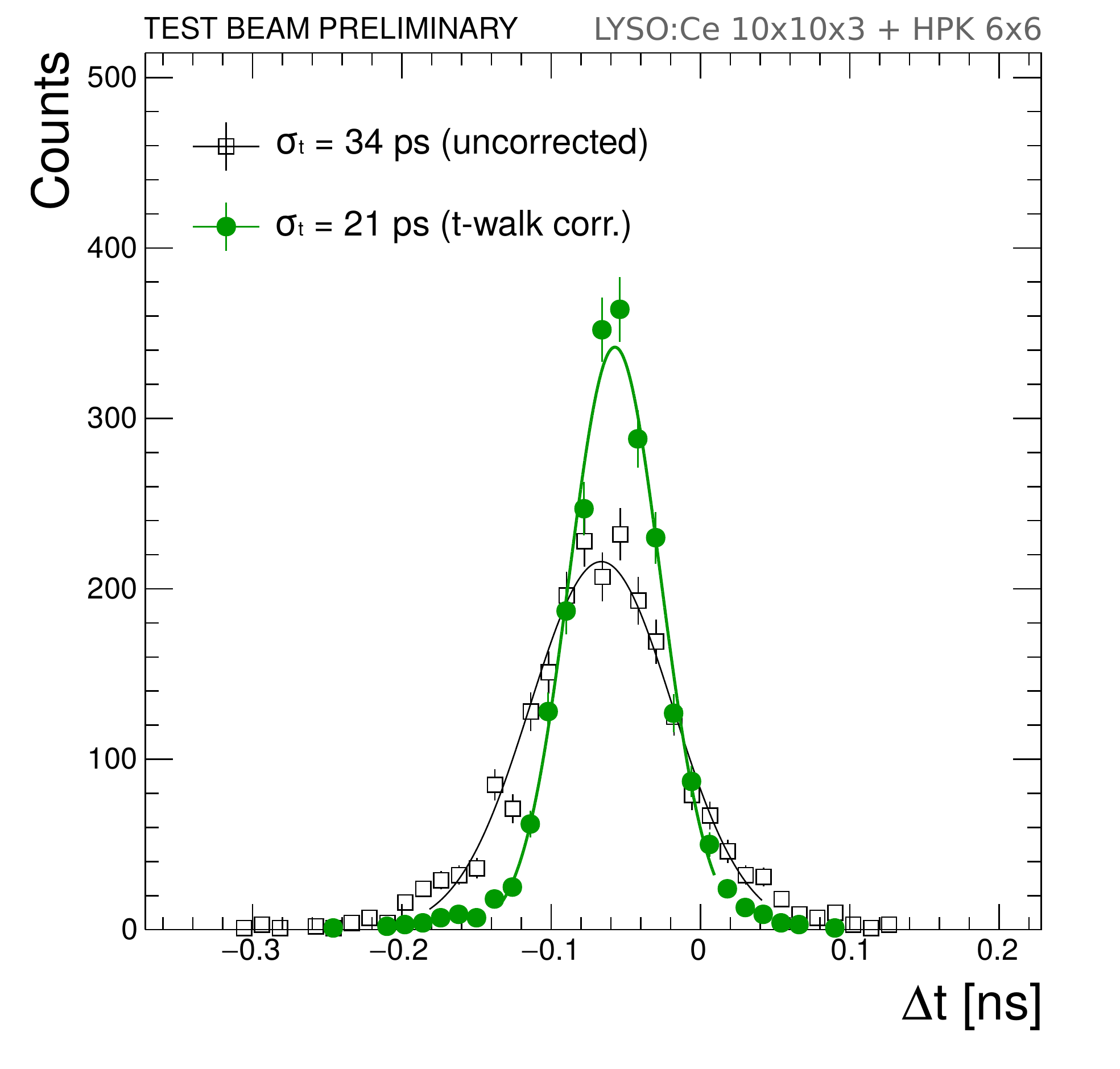}}
\raisebox{-0.5\height}{\includegraphics[width=0.52\textwidth]{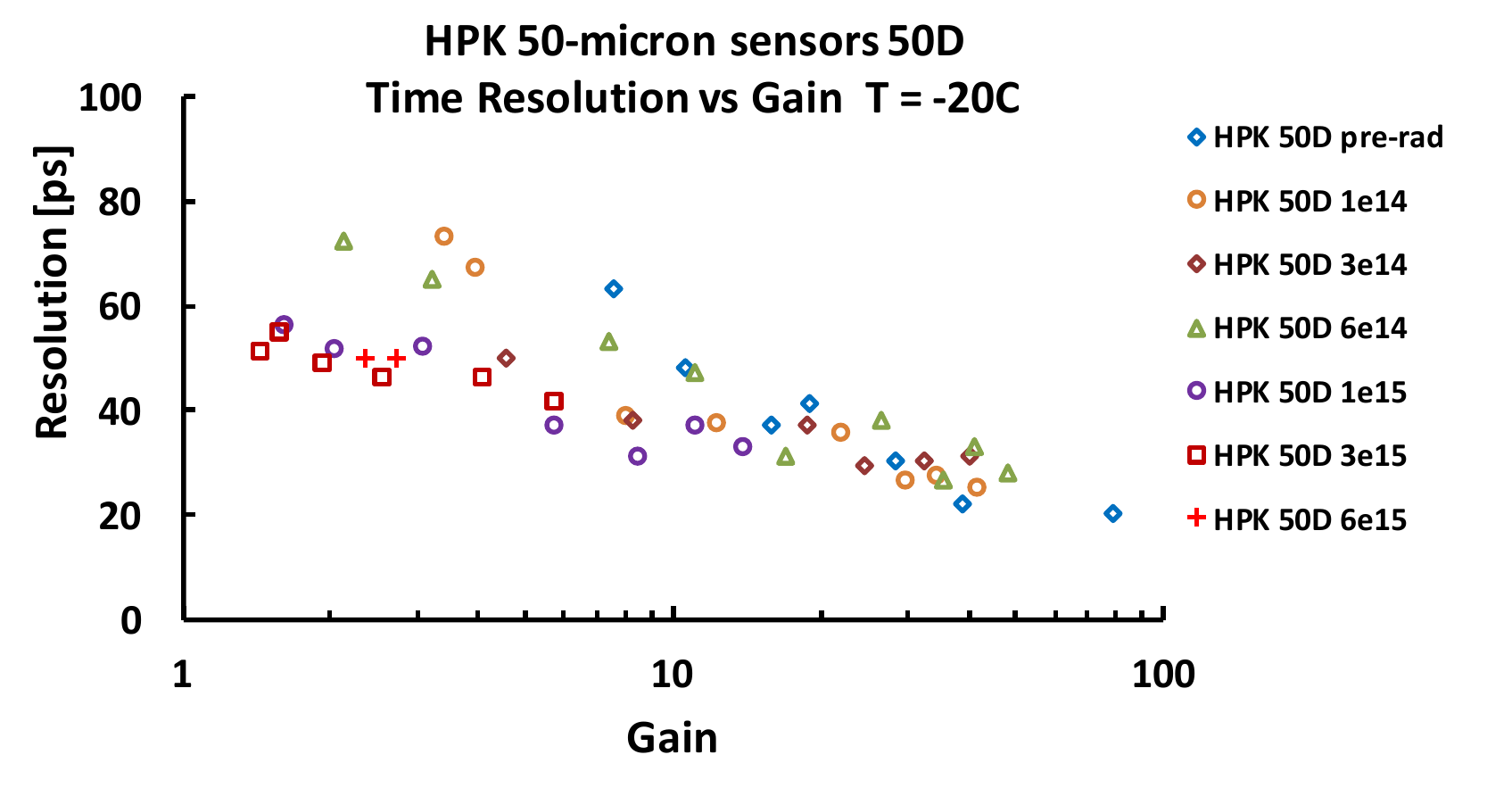}}
\caption{Distribution of the time difference in a pair of LYSO:Ce tiles exposed to a 3 mm wide beam of MIPs hitting the centre of the tiles. Results before and after time walk correction for $10\times10\times3$ mm$^{3}$ crystals read out with $6\times6$ mm$^{2}$ HPK SiPMs  are shown (left).  Time resolution for HPK LGAD sensors as a function of internal gain, different radiation levels are shown with different markers (right). Relevant gains for CMS are $\sim$\hspace{-0.03cm}$10$. }
\label{fig:timeres}
\end{figure}

The barrel timing layer will cover the pseudorapidity region up to $|\eta| = 1.48$ with a total active surface of about 40~m$^2$. The  fundamental detection cell unit will consist of a thin  $\sim12\times12$~mm$^2$ LYSO:Ce crystal coupled to a $4\times4$~mm$^2$ SiPM. The crystal thickness will vary between about 3.7~mm ($|\eta|<0.7$) and 2.4~mm ($|\eta|>1.1$), to equalize the slant depth crossed by all particles starting from the interaction point. LYSO based scintillators and SiPM devices are mature technologies, with production and assembly procedures well established and standardized in industry.
The BTL will share CO$_{2}$ cooling with the tracker. Trays consisting of LYSO:Ce+SiPM modules will be inserted into the Tracker Support Tube (TST). Small prototypes consisting of LYSO:Ce crystals read out with SiPMs have been proven capable of achieving time resolution below 30~ps (see Fig.~\ref{fig:timeres})~\cite{Benaglia:2016eya} in various testbeams that were held at CERN and Fermilab.
Both the crystals and the SiPM are proven to be radiation tolerant up to a neutron equivalent fluence of $2\times10^{14}$~cm$^{-2}$, when cooled to below $-30$~$^{\circ}$C.
The read-out electronics are based on existing positron emission tomography applications with time-of-flight (TOF-PET) measurement~\cite{Rolo2013-ASIC, Rolo2011-CMOS, DiFrancesco2016-TOFPET2}, which will be readout every time there is an external (Level-1) trigger.

The proposed layout has no significant impact on the upgraded designs, performances and schedules of the Tracker and the ECAL.

\subsection{Endcap Timing Layer}

We propose to  build the fast-timing detector in the forward regions ($|\eta|=1.6$ to $|\eta|=2.9$) based on Low-Gain-Avalanche-Diodes (LGAD)~\cite{Pellegrini201412, Cartiglia:2015iua}, which are also considered for a fast-timing layer in the very forward region ($2.4 < |\eta| < 4.8$) of the ATLAS experiment~\cite{CERN-LHCC-2015-020}. The technology selected for the barrel cannot be extended to the endcap, due to radiation tolerance limitations. The radiation tolerance studies for LGADs, though still in progress, indicate the promising performance of about 30 and 50~ps at fluences corresponding to $|\eta|\simeq2.5$ and $3.0$, respectively, at the end of the HL-LHC operation (see Fig.~\ref{fig:timeres}).
Achieving good time performance at low-gain requires cell sizes typically less than 3~mm$^2$, to limit the sensor capacitance.

We propose to place the endcap timing layer (ETL) in a thermally isolated volume on the nose of the endcap electromagnetic calorimeter, at a distance of about 3~m from the interaction point, to ensure its accessibility throughout the HL-LHC operation. Sensors will be arranged on both sides of the discs two ensure full coverage. The active sensor area is a little over 6~m$^{2}$ per endcap disk.

While the barrel timing layer should be installed before the Tracker upgrade, the schedule to assemble the endcap disks can extend until the end of the LHC long shutdown LS3 (2025), thus providing additional time to complete the R\&D plan for the sensors.

\section{Summary and Future Considerations}
The addition of a timing layer will allow us to suppress the effect of pileup on all object-level observables at the HL-LHC. This will contribute to improvements in many physics analyses -- including Higgs, BSM, LLPs etc. --  by increasing signal efficiencies or reducing the width of residual distributions for discriminating variables.

 The MTD is composed of two subsystems based on different sensor technologies:  SiPMs+LYSO:Ce for the barrel region and LGADs for the endcap region. The technology choices were driven by performance, radiation, mechanics and schedule requirements and constraints. Both sensor technologies are mature and proven to achieve $\sim$\hspace{-0.01cm}30--50~ps time resolutions.

Another interesting development will be the impact of timing on the CMS L1 triggers. The technical implications for the MTD read-out electronics for the Level-1 trigger latency, as well as a complete description of the mechanical specifications and detector performance will be described in the  forthcoming Technical Design Report (TDR) from CMS.


\bibliography{eprint}

\begin{thebibliography}{10}

\bibitem{Apollinari:2017cqg}
G.~Apollinari, O.~Brüning, T.~Nakamoto, and L.~Rossi, ``{High Luminosity Large
  Hadron Collider HL-LHC},'' {\em CERN Yellow Report}, p.~1, 2015.

\bibitem{CMS:2009nxa}
``Particle-flow reconstruction and global event description with the cms
  detector,'' {\em JINST}, vol.~12, p.~P10003, 2017.

\bibitem{Bertolini:2014bba}
D.~Bertolini, P.~Harris, M.~Low, and N.~Tran, ``{Pileup Per Particle
  Identification},'' {\em JHEP}, vol.~1410, p.~59, 2014.

\bibitem{Butler:2020886}
``{Technical Proposal for the Phase-II Upgrade of the CMS Detector},'' Tech.
  Rep. CERN-LHCC-2015-010. LHCC-P-008, CERN, Geneva, Jun 2015.

\bibitem{Butler:2055167}
``{CMS Phase II Upgrade Scope Document},'' Tech. Rep. CERN-LHCC-2015-019.
  LHCC-G-165, CERN, Geneva, Sep 2015.

\bibitem{Collaboration:2296612}
C.~Collaboration, ``{TECHNICAL PROPOSAL FOR A MIP TIMING DETECTOR IN THE CMS
  EXPERIMENT PHASE 2 UPGRADE},'' Tech. Rep. CERN-LHCC-2017-027. LHCC-P-009,
  CERN, Geneva, Dec 2017.
\newblock This document describes a MIP timing detector for the Phase-2 upgrade
  of the CMS experiment, in view of HL-LHC running.

\bibitem{gundacker2013time}
S.~Gundacker, E.~Auffray, B.~Frisch, P.~Jarron, A.~Knapitsch, T.~Meyer,
  M.~Pizzichemi, and P.~Lecoq, ``{Time of flight positron emission tomography
  towards 100~ps resolution with L(Y)SO: an experimental and theoretical
  analysis},'' {\em JINST}, vol.~8, p.~P07014, 2013.

\bibitem{LYSONIM}
D.~Anderson, A.~Apresyan, A.~Bornheim, J.~Duarte, C.~Pena, A.~Ronzhin,
  M.~Spiropulu, J.~Trevor, and S.~Xie, ``On timing properties of lyso-based
  calorimeters,'' {\em Nucl. Instrum. Meth. A}, vol.~794, p.~7, 2015.

\bibitem{Anderson:2015tia}
D.~Anderson {\em et~al.}, ``{Precision Timing Measurements for High Energy
  Photons},'' {\em Nucl.Instrum.Meth. A}, vol.~787, p.~94, 2015.

\bibitem{White:2014oga}
S.~N. White, ``{R\&D for a Dedicated Fast Timing Layer in the CMS Endcap
  Upgrade},'' {\em Acta Phys. Pol. B Proc. Suppl.}, vol.~7, p.~743, 2014.

\bibitem{Pellegrini201412}
G.~Pellegrini {\em et~al.}, ``{Technology developments and first measurements
  of Low Gain Avalanche Detectors (LGAD) for high energy physics
  applications},'' {\em Nucl. Instrum. Meth. A}, vol.~765, p.~12, 2014.

\bibitem{Cartiglia:2015iua}
N.~Cartiglia {\em et~al.}, ``{Design optimization of ultra-fast silicon
  detectors},'' {\em Nucl. Instrum. Meth. A}, vol.~796, p.~141, 2015.

\bibitem{Benaglia:2016eya}
A.~Benaglia, S.~Gundacker, P.~Lecoq, M.~T. Lucchini, A.~Para, K.~Pauwels, and
  E.~Auffray, ``{Detection of high energy muons with sub-20 ps timing
  resolution using L(Y)SO crystals and SiPM readout},'' {\em Nucl. Instrum.
  Meth. A}, vol.~830, p.~30, 2016.

\bibitem{Rolo2013-ASIC}
M.~D. Rolo {\em et~al.}, ``{TOFPET} {ASIC} for {PET} applications,'' {\em
  JINST}, vol.~8, p.~C02050, 2013.

\bibitem{Rolo2011-CMOS}
M.~D. Rolo {\em et~al.}, ``A low-noise {CMOS} front-end for {TOF}-{PET},'' {\em
  JINST}, vol.~6, p.~P09003, 2011.

\bibitem{DiFrancesco2016-TOFPET2}
A.~D. Francesco {\em et~al.}, ``{TOFPET2}: a high-performance {ASIC} for time
  and amplitude measurements of {SiPM} signals in time-of-flight
  applications,'' {\em JINST}, vol.~11, p.~C03042, 2016.

\bibitem{CERN-LHCC-2015-020}
``{ATLAS Phase-II Upgrade Scoping Document},'' Tech. Rep. CERN-LHCC-2015-020.
  LHCC-G-166, CERN, Geneva, Sep 2015.

\end{thebibliography}
\bibliographystyle{ieeetr}

\end{document}